\documentclass[prb,superscriptaddress,amsfonts,amsmath,floatfix]{revtex4}
\usepackage{graphicx}
\usepackage{bm}
\usepackage{amsmath}
\usepackage{hyperref}

\begin{document}
\title{Topological and Hopf charges of a twisted Skyrmion string}

\author{Malcolm Anderson}
\email{malcolm.anderson@ubd.edu.bn}
\affiliation{Department of Mathematics, Universiti Brunei Darussalam\\
Jalan Tungku Link BE1410, Gadong, Negara Brunei Darussalam}
\author{Miftachul Hadi}
\email{itpm.id@gmail.com}
\affiliation{Department of Mathematics, Universiti Brunei Darussalam\\
Jalan Tungku Link BE1410, Gadong, Negara Brunei Darussalam}
\affiliation{Physics Research Centre, Indonesian Insitute of Sciences (LIPI)\\
Kompleks Puspiptek, Serpong, Tangerang 15314, Indonesia}
\affiliation{Department of Physics, School of Natural Sciences\\
Ulsan National Institute of Science and Technology (UNIST)\\
50, UNIST-gil, Eonyang-eup, Ulju-gun, Ulsan, South Korea}
\affiliation{Institute of Modern Physics, Chinese Academy of Sciences\\
509 Nanchang Rd., Lanzhou  730000, China}

\date{\today}
\begin{abstract}
We study topological and Hopf charges of a twisted Skyrmion string. Topological and Hopf charges are defined with the same definition, only differ in vortex solution component. 
\end{abstract}

\maketitle

\section{BACKGROUND}
On Jul 7, 2015 M. Nitta mentioned that: In some of recent your papers I found you discuss twisted baby Skyrmion strings. We also studied the same soliton on $R^2\times S^1$ (http://arxiv.org/abs/1305.7417) \cite{nitta1}. In addition to the usual topological charge $\pi_2$ for baby Skyrmions, these objects carry additional topological charge related to the Hopf charge $\pi_3$. Exactly speaking in our case, it is a mathematically different charge because of a compactified geometry. I think that your objects carry the same charge \cite{nitta2}.

Nitta has kindly pointed out that, for twisting solutions like the twisted baby Skyrmion string, there is a second conserved quantity (the Hopf charge) in addition to the topological charge. For our twisted solutions, which depend on $n\theta+mkz$, the Hopf charge is actually proportional to $nmk$. But since the topological charge $n$ is conserved, it follows that conservation of the Hopf charge is equivalent to the conservation of $mk$. The question of whether $mk$ is conserved is part of a much more general question of whether our self-gravitating string solutions are stable, and this is something well beyond where we are at the moment. Calculating the value of the Hopf charge for a twisted Skyrmion string i.e. for an infinite string, the answer is obviously undefined anyway (as $\Delta z = \infty$). In the other words, the Hopf charge also diverges if we integrate over all $z$ from $-\infty$ to $\infty$, which is why it really only makes sense for compact solutions (like Nitta's) for which the range of $z$ is finite.

Geometrically, the Hopf charge measures the number of times the solution twists a full circle over its length in the $z$-direction. The fact that it is conserved means that if the solution is perturbed then it will still twist the same number of turns over its length, no matter how it is distorted. But of course for our twisted vortex solutions, the total number of twists is infinite because the length of the string is infinite, so a more useful idea in this case is that the average number of twists per unit length is conserved, which is to say that $mk$ is constant.

If $mk$ is conserved for the self-gravitating strings, this does not necessarily mean that they are stable. There are many ways they could be unstable: they could collapse inwards to form a line with infinite density, or they could expand outwards. However, it is also possible that they might gravitationally "radiate away" the twists (much as a cosmic string which is almost straight but has small "bumps" is believed to radiate the energy in the bumps away). But, we have no idea and it is not possible to talk about the stability of the solutions unless we first find some solutions.

\section{TOPOLOGICAL CHARGE OF A TWISTED SKYRMION STRING}
Let us discuss topological charge in more detail. Here, topological charge is denoted by
\begin{equation}
T=\frac{1}{4\pi}~\varepsilon^{abc}\int\int_A \phi_a~\frac{\partial\phi_b}{\partial x}~\frac{\partial\phi_c}{\partial y}~dx~dy
\end{equation}
where $A$ is any plane parallel to the $x-y$ plane, and 
\begin{equation}
\varepsilon^{abc}=
\left\{
\begin{array}{ll}
1,  & \text{if}~\left\{a,b,c\right\}~\text{is an even permutation of}~\left\{1,2,3\right\}\\
-1, & \text{if}~\left\{a,b,c\right\}~\text{is an odd permutation of}~\left\{1,2,3\right\}\\
0,  & \text{if}~a=b,~\text{or}~b=c,~\text{or}~c=a.
\end{array}\right.
\end{equation}
This is Levi-Civita symbol in three dimensions.
So,
\begin{equation}
\varepsilon^{123}=\varepsilon^{231}=\varepsilon^{312}=1;~~~\varepsilon^{213}=\varepsilon^{132}=\varepsilon^{321}=-1
\end{equation}
where all others, e.g. $\varepsilon^{111}$, $\varepsilon^{122}$, $\varepsilon^{323}$ are zero.

For the vortex solution, we use ansatz as below
\begin{equation}
\phi_a
=\begin{pmatrix}
\sin f(r)~\sin(n\theta-\chi)\\
\sin f(r)~\cos(n\theta-\chi)\\
\cos f(r)
\end{pmatrix}
\end{equation}
To simplify calculation, let us assume that
\begin{eqnarray}
\sin f(r)=s;~~~\cos f(r)=c;~~~\sin(n\theta-\chi)=S;~~~\cos(n\theta-\chi)=C
\end{eqnarray}
then vortex solution can be writen as
\begin{equation}
\phi_a=
\begin{pmatrix}
\sin f(r)~\sin(n\theta-\chi)\\
\sin f(r)~\cos(n\theta-\chi)\\
\cos f(r)
\end{pmatrix}
=
\begin{pmatrix}
sS\\
sC\\
c
\end{pmatrix}
=
\begin{pmatrix}
\phi_1\\
\phi_2\\
\phi_3 
\end{pmatrix}
\end{equation}
From (1), (6), we obtain
\begin{eqnarray}
\varepsilon^{abc}~\phi_a~\frac{\partial\phi_b}{\partial x}~\frac{\partial\phi_c}{\partial y}
&=&\varepsilon^{123}~\phi_1~\frac{\partial\phi_2}{\partial x}~\frac{\partial\phi_3}{\partial y}+\varepsilon^{231}~\phi_2~\frac{\partial\phi_3}{\partial x}~\frac{\partial\phi_1}{\partial y}+~\varepsilon^{312}~\phi_3~\frac{\partial\phi_1}{\partial x}~\frac{\partial\phi_2}{\partial y}+\varepsilon^{213}~\phi_2~\frac{\partial\phi_1}{\partial x}~\frac{\partial\phi_3}{\partial y}\nonumber\\
&&+~\varepsilon^{132}\phi_1~\frac{\partial\phi_3}{\partial x}~\frac{\partial\phi_2}{\partial y}+\varepsilon^{321}\phi_3~\frac{\partial\phi_2}{\partial x}~\frac{\partial\phi_1}{\partial y}
\end{eqnarray}

\begin{eqnarray}
\frac{\partial\phi_a}{\partial x}
&=&
\frac{\partial}{\partial x}
\begin{pmatrix}
\sin f(r)~ \sin(n\theta-\chi) \\
\sin f(r)~ \cos(n\theta-\chi) \\
\cos f(r)
\end{pmatrix}=
\begin{pmatrix}
c\frac{\partial f}{\partial x}S+sC~n~\frac{\partial\theta}{\partial x}\\
c~\frac{\partial f}{\partial x}~C-sS~n~\frac{\partial\theta}{\partial x}\\
-s~\frac{\partial f}{\partial x}
\end{pmatrix}=
\begin{pmatrix}
cS\\
cC\\
-s
\end{pmatrix}
\frac{\partial f}{\partial x}+n
\begin{pmatrix}
sC\\
-sS\\
0
\end{pmatrix}
\frac{\partial\theta}{\partial x}\nonumber\\
&=& W_a\frac{\partial f}{\partial x}+nV_a
\frac{\partial\theta}{\partial x}
\end{eqnarray}

\begin{eqnarray}
\frac{\partial\phi_a}{\partial y}
&=&
\begin{pmatrix}
cS\\
cC\\
-s
\end{pmatrix}
\frac{\partial f}{\partial y}+n
\begin{pmatrix}
sC\\
-sS\\
0
\end{pmatrix}
\frac{\partial\theta}{\partial y}=
W_a\frac{\partial f}{\partial y}+nV_a\frac{\partial\theta}{\partial y}
\end{eqnarray}
where
\begin{equation}
W_a=
\begin{pmatrix}
cS\\
cC\\
-s
\end{pmatrix};~~~
V_a=
\begin{pmatrix}
sC\\
-sS\\
0
\end{pmatrix}
\end{equation}
Substitute (9) into (7), we obtain
\begin{eqnarray}
\varepsilon^{abc}~\phi_a~\frac{\partial\phi_b}{\partial x}~\frac{\partial\phi_c}{\partial y}
&=&\varepsilon^{abc}~\phi_a~\left[W_b\frac{\partial f}{\partial x}+nV_b
\frac{\partial\theta}{\partial x}\right]\left[W_c\frac{\partial f}{\partial y}+nV_c
\frac{\partial\theta}{\partial y}\right]\nonumber\\
&=&\varepsilon^{abc}~\phi_a~W_bW_c\frac{\partial f}{\partial x}\frac{\partial f}{\partial y}+\varepsilon^{abc}~\phi_a~W_bV_cn\frac{\partial f}{\partial x}\frac{\partial\theta}{\partial y}
+\varepsilon^{abc}~\phi_a~V_bW_cn\frac{\partial\theta}{\partial x}\frac{\partial f}{\partial y}\nonumber\\
&&+~\varepsilon^{abc}~\phi_a~V_bV_cn^2\frac{\partial\theta}{\partial x}\frac{\partial\theta}{\partial y}
\end{eqnarray}
Here
\begin{equation}
\varepsilon^{abc}\phi_a~V_b~V_c=0
\end{equation}
because (e.g.)
\begin{eqnarray}
\varepsilon^{1bc}~\phi_1~V_b~V_c
&=&\varepsilon^{123}~\phi_1~V_2~V_3+\varepsilon^{132}~\phi_1~V_3~V_2=\phi_1~V_2~V_3-\phi_1~V_3~V_2=0.
\end{eqnarray}
The same is true with $a=2$ and with $a=3$.

Similarly
\begin{equation}
\varepsilon^{abc}~\phi_a~W_1~W_c=0
\end{equation}
So,
\begin{equation}
\varepsilon^{abc}~\phi_a~\frac{\partial\phi_b}{\partial x}~\frac{\partial\phi_c}{\partial y}
=\varepsilon^{abc}~\phi_a~W_b~V_c~n~\frac{\partial f}{\partial x}~\frac{\partial\theta}{\partial y}+\varepsilon^{abc}~\phi_a~V_b~W_c~n~\frac{\partial\theta}{\partial x}~\frac{\partial f}{\partial y}
\end{equation}
because
\begin{eqnarray}
\varepsilon^{abc}~\phi_a~W_b~V_c
&=&\varepsilon^{acb}~\phi_a~W_c~V_b=-\varepsilon^{abc}~\phi_a~V_b~W_c.
\end{eqnarray}
(Note: the triple scalar product of vectors is antisymmetric when exchanging any pair of arguments). For example: 
\begin{equation}
\varepsilon_{ijk}~a^i~b^j~c^k=\mathbf{a}~.~(\mathbf{b}\times\mathbf{c})=\mathbf{b}~.~(\mathbf{c}\times\mathbf{a})=-\mathbf{b}~.~(\mathbf{a}\times\mathbf{c})
\end{equation}
Therefore,
\begin{eqnarray}
\varepsilon^{abc}~\phi_a~\frac{\partial\phi_b}{\partial x}~\frac{\partial\phi_c}{\partial y}
&=&-\varepsilon^{abc}~\phi_a~V_b~W_c~n~\frac{\partial f}{\partial x}~\frac{\partial\theta}{\partial y}+\varepsilon^{abc}~\phi_a~V_b~W_c~n~\frac{\partial\theta}{\partial x}~\frac{\partial f}{\partial y}\nonumber\\
&=&n~\varepsilon^{abc}~\phi_a~V_b~W_c~\left(\frac{\partial\theta}{\partial x}\frac{\partial f}{\partial y}-\frac{\partial f}{\partial x}\frac{\partial\theta}{\partial y}\right)
\end{eqnarray}
Here,
\begin{eqnarray}
\varepsilon^{abc}~\phi_a~V_b~W_c
&=&\varepsilon^{123}~\phi_1~V_2~W_3+\varepsilon^{231}~\phi_2~V_3~W_1+\varepsilon^{312}~\phi_3~V_1~W_2+\varepsilon^{213}~\phi_2~V_1~W_3=s.
\end{eqnarray}
where
\begin{eqnarray}
W_a=
\begin{pmatrix}
cS\\
cC\\
-s
\end{pmatrix}
=
\begin{pmatrix}
W_1\\
W_2\\
W_3
\end{pmatrix}
;~~~
V_a=
\begin{pmatrix}
sC\\
-sS\\
0
\end{pmatrix}
=
\begin{pmatrix}
V_1\\
V_2\\
V_3
\end{pmatrix}
\end{eqnarray}

\begin{equation}
\underline{\phi}~.~(\underline{V}\times\underline{W})=\phi_1~(V_2W_3-V_3W_2)+\phi_2~(V_3W_1-V_1W_3)+\phi_3~(V_1W_2-V_2W_1)
\end{equation}
Hence,
\begin{equation}
\varepsilon^{abc}~\phi_a~\frac{\partial\phi_b}{\partial x}~\frac{\partial\phi_c}{\partial y}=ns~\left(\frac{\partial\theta}{\partial x}\frac{\partial f}{\partial y}-\frac{\partial f}{\partial x}\frac{\partial\theta}{\partial y}\right)
\end{equation}
For the vortex solution
\begin{equation}
f=f(r)
\end{equation}
Here, we use cylindrical coordinates $(r,\theta)$, where $r=(x^2+y^2)^{1/2}$ and $\theta=\tan^{-1}\left(\frac{x}{y}\right)$. So, we obtain
\begin{eqnarray}
\frac{\partial f}{\partial x}
&=&\frac{\partial f}{\partial r}\frac{\partial r}{\partial x} +\frac{\partial f}{\partial\theta}\frac{\partial\theta}{\partial x}  = \frac{1}{r}\frac{\partial f}{\partial r}~x+\frac{1}{r^2}\frac{\partial f}{\partial\theta}~y
\end{eqnarray}

\begin{eqnarray}
\frac{\partial f}{\partial y}
&=&\frac{\partial f}{\partial r}\frac{\partial r}{\partial y} +\frac{\partial f}{\partial\theta}\frac{\partial\theta}{\partial y}  = \frac{1}{r}\frac{\partial f}{\partial r}~y-\frac{1}{r^2}\frac{\partial f}{\partial\theta}~x = \frac{1}{r}\frac{\partial f}{\partial r}~y
\end{eqnarray}
due to $f$ is a function of $r$ only. 

So,
\begin{equation}
\frac{\partial f}{\partial x}=\frac{x}{r}f';~~~
\frac{\partial f}{\partial y}=\frac{y}{r}f'
\end{equation}
Using relation below
\begin{eqnarray}
\frac{\partial}{\partial x}\tan^{-1}u
&=& \frac{1}{1+u^2}~\frac{du}{dx}
\end{eqnarray}
for $-\frac{\pi}{2}<\tan^{-1}u<\frac{\pi}{2}$. Also
\begin{eqnarray}
\frac{d}{dx}\left(\frac{u}{v}\right)
&=& \frac{u'~v-u~v'}{v^2}
\end{eqnarray}
Then we obtain
\begin{equation}
\frac{\partial\theta}{\partial x}=\frac{y}{x^2+y^2}=\frac{y}{r^2};~~~\frac{\partial\theta}{\partial y}=-\frac{x}{x^2+y^2}=-\frac{x}{r^2}
\end{equation}
So,
\begin{eqnarray}
\frac{\partial\theta}{\partial x}~\frac{\partial f}{\partial y}-\frac{\partial f}{\partial x}\frac{\partial\theta}{\partial y}
&=&\frac{y}{r^2}~\frac{y}{r}f'-\frac{x}{r}f'~\left(-\frac{x}{r^2}\right)=\frac{1}{r}~f'.
\end{eqnarray}
Hence,
\begin{equation}
\varepsilon^{abc}~\phi_a~\frac{\partial\phi_b}{\partial x}~\frac{\partial\phi_c}{\partial y}=ns~\left(\frac{\partial\theta}{\partial x}\frac{\partial f}{\partial y}-\frac{\partial f}{\partial x}\frac{\partial\theta}{\partial y}\right)=n~\sin f~\frac{1}{r}f'
\end{equation}
Now, we want to integrate
\begin{eqnarray}
T
&=&\frac{1}{4\pi}~\varepsilon^{abc}\int\int~\phi_a~\frac{\partial\phi_b}{\partial x}~\frac{\partial\phi_c}{\partial y}~dx~dy=\frac{1}{4\pi}\int\int~n~\sin f~\frac{1}{r}~f'~dx~dy.
\end{eqnarray}
but
\begin{equation}
dx~dy=r~dr~d\theta
\end{equation}
So,
\begin{eqnarray}
T
&=&\frac{1}{4\pi}\int\int~n~\sin f~\frac{1}{r}~f'~r~dr~d\theta=\frac{n}{4\pi}\int_0^\infty\sin f~f'~dr\int_0^{2\pi} d\theta\nonumber\\
&=&\frac{n}{4\pi}\int_0^\infty\sin f~f'~dr~\times 2\pi=\frac{n}{2}\int_0^\infty\sin f~f'~dr\nonumber\\
&=&\frac{n}{2}\left.(-\cos f)\right|_{0}^\infty
\end{eqnarray}
Because (Chain Rule),
\begin{equation}
\frac{d}{dr}\cos f(r)=\left(\frac{d}{df}~\cos f\right)~\frac{df}{dr}=-\sin f~f'
\end{equation}
So,
\begin{eqnarray}
T
&=&\frac{n}{2}\left.[-\cos f(r)]\right|_{r=0}^{r=\infty}=-\frac{n}{2}[\cos f(\infty)-\cos f(0)]=-\frac{n}{2}[\cos 0-\cos\pi]=-\frac{n}{2}[1-(-1)]=-\frac{n}{2}\times[2]\nonumber\\
&=&-n.
\end{eqnarray}
where $n$ is winding number and we use boundary conditions for vortex, i.e. 
\begin{eqnarray}
\lim_{r\to \infty}f(r)=0;~~~
\lim_{r\to 0}f(r)=\pi.
\end{eqnarray}
It means that for the vortex solution, the topological charge is just the winding number, $n$.

We can show that the topological charge is conserved. That is
\begin{equation}
\frac{dT}{dt}=0
\end{equation}
no matter what solution $\phi_a$ we have.

Let us write the topological charge as below
\begin{equation}
T=\frac{1}{4\pi}~\varepsilon^{abc}\int\int \phi_a~\frac{\partial\phi_b}{\partial x}~\frac{\partial\phi_c}{\partial y}~dx~dy
\end{equation}
So,
\begin{eqnarray}
\frac{dT}{dt}
&=&\frac{1}{4\pi}~\varepsilon^{abc}\int\int\frac{\partial\phi_a}{\partial t}~\frac{\partial\phi_b}{\partial x}~\frac{\partial\phi_c}{\partial y}~dx~dy+\frac{1}{4\pi}~\varepsilon^{abc}\int\int\phi_a~\frac{\partial^2\phi_b}{\partial x\partial t}~\frac{\partial\phi_c}{\partial y}~dx~dy\nonumber\\
&&+~\frac{1}{4\pi}~\varepsilon^{abc}\int\int\phi_a~\frac{\partial\phi_b}{\partial x}~\frac{\partial^2\phi_c}{\partial y\partial t}~dx~dy
\end{eqnarray}
Gauss' theorem gives
\begin{eqnarray}
\int\int\phi_a~\frac{\partial^2\phi_b}{\partial x\partial t}~\frac{\partial\phi_c}{\partial y}~dx~dy
&=&\left[\int\phi_a~\left.\frac{\partial\phi_b}{\partial t}~\frac{\partial\phi_c}{\partial y}~dy\right]\right|_{x=-\infty}^\infty-\int\int\frac{\partial\phi_b}{\partial t}~\frac{\partial}{\partial x}\left[\phi_a~\frac{\partial\phi_c}{\partial y}\right]~dx~dy\nonumber\\
&=&-\int\int\frac{\partial\phi_b}{\partial t}\left(\frac{\partial\phi_a}{\partial x}~\frac{\partial\phi_c}{\partial y}+\phi_a~\frac{\partial^2\phi_c}{\partial x\partial y}\right)~dx~dy
\end{eqnarray}
Boundary term is zero i.e:
\begin{equation}
\frac{\partial\phi_c}{\partial y}\rightarrow 0,~~~\text{at}~~\infty
\end{equation}
Similarly,
\begin{equation}
\int\int\phi_a~\frac{\partial\phi_b}{\partial x}~\frac{\partial^2\phi_c}{\partial y\partial t}~dx~dy=-\int\int\frac{\partial\phi_c}{\partial t}~\left(\frac{\partial\phi_a}{\partial y}\frac{\partial\phi_b}{\partial x}+\phi_a~\frac{\partial^2\phi_b}{\partial x\partial y}\right)~dx~dy
\end{equation}
So,
\begin{eqnarray}
\int\int\phi_a~\frac{\partial^2\phi_b}{\partial x\partial t}~\frac{\partial\phi_c}{\partial y}~dx~dy
&+&~\int\int \phi_a~\frac{\partial\phi_b}{\partial x}~\frac{\partial^2\phi_c}{\partial y\partial t}~dx~dy\nonumber\\
&=&-\int\int\frac{\partial\phi_b}{\partial t}\left(\frac{\partial\phi_a}{\partial x}~\frac{\partial\phi_c}{\partial y}+\phi_a~\frac{\partial^2\phi_c}{\partial x\partial y}\right)~dx~dy\nonumber\\
&&-~\int\int\frac{\partial\phi_c}{\partial t}~\left(\frac{\partial\phi_a}{\partial y}\frac{\partial\phi_b}{\partial x}+\phi_a~\frac{\partial^2\phi_b}{\partial x\partial y}\right)~dx~dy\nonumber\\
&=&-\int\int \left(\frac{\partial\phi_b}{\partial t}~\frac{\partial\phi_a}{\partial x}~\frac{\partial\phi_c}{\partial y}+\frac{\partial\phi_c}{\partial t}~\frac{\partial\phi_a}{\partial y}~\frac{\partial\phi_b}{\partial x}\right)~dx~dy\nonumber\\
&&-\int\int\left(\frac{\partial\phi_b}{\partial t}~\phi_a~\frac{\partial^2\phi_c}{\partial x\partial y}+\frac{\partial\phi_c}{\partial t}~\phi_a~\frac{\partial^2\phi_b}{\partial x\partial y}\right)dxdy.
\end{eqnarray}
Because $(bac)$ is odd/even when $(cab)$ is even/odd, we find that
\begin{eqnarray}
\varepsilon^{abc}\int\int\phi_a~\frac{\partial^2\phi_b}{\partial x\partial t}~\frac{\partial\phi_c}{\partial y}~dx~dy
&+&~\varepsilon^{abc}\int\int\phi_a~\frac{\partial\phi_b}{\partial x}~\frac{\partial^2\phi_c}{\partial y\partial t}~dx~dy\nonumber\\
&-&\varepsilon^{abc}\int\int\left(\frac{\partial\phi_b}{\partial t}~\frac{\partial\phi_a}{\partial x}~\frac{\partial\phi_c}
{\partial y}+\frac{\partial\phi_c}{\partial t}~\frac{\partial\phi_a}{\partial y}~\frac{\partial\phi_b}{\partial x}\right)~dx~dy\nonumber\\
&-&\varepsilon^{abc}\int\int\left(\frac{\partial\phi_b}{\partial t}~\phi_a~\frac{\partial^2\phi_c}{\partial x\partial y}+\frac{\partial\phi_c}{\partial t}~\phi_a~\frac{\partial^2\phi_b}{\partial x\partial y}\right)~dx~dy\nonumber\\
&=&\int\int 0~dx~dy+\int\int 0~dx~dy\nonumber\\
&=&0
\end{eqnarray}
Finally, we see that
\begin{equation}
\frac{dT}{dt}=\frac{1}{4\pi}~\varepsilon^{abc}\int\int\frac{\partial\phi_a}{\partial t}~\frac{\partial\phi_b}{\partial x}~\frac{\partial\phi_c}{\partial y}~dx~dy
\end{equation}
So,
\begin{eqnarray}
\frac{\partial\phi_a}{\partial t}
&=&
\begin{pmatrix}
cS\\
cC\\
-s
\end{pmatrix}
\frac{\partial f}{\partial t}+
\begin{pmatrix}
sC\\
-sS\\
0
\end{pmatrix}
\frac{\partial g}{\partial t}=W_a~\frac{\partial f}{\partial t}+V_a~\frac{\partial g}{\partial t}.
\end{eqnarray}
Similarly,
\begin{equation}
\frac{\partial \phi_b}{\partial x}=W_b~\frac{\partial f}{\partial x}+V_b~\frac{\partial g}{\partial x};~~~\frac{\partial \phi_c}{\partial y}=W_c~\frac{\partial f}{\partial y}+V_c~\frac{\partial g}{\partial y}
\end{equation}
So,
\begin{eqnarray}
\varepsilon^{abc}~\frac{\partial\phi_a}{\partial t}~\frac{\partial\phi_b}{\partial x}~\frac{\partial\phi_c}{\partial y}
&=&\varepsilon^{abc}\left(V_a~\frac{\partial g}{\partial t}+W_a~\frac{\partial f}{\partial t}\right)\left(V_b~\frac{\partial g}{\partial x}+W_b~\frac{\partial f}{\partial x}\right)\left(V_c~\frac{\partial g}{\partial y}+W_c~\frac{\partial f}{\partial y}\right)= 0.
\end{eqnarray}
But, refer to triple scalar products of vector using Levi-Civita symbol, we obtain
\begin{eqnarray}
\varepsilon^{abc}~V_a~W_b~X_c
=\underline{V}~.~(\underline{W}\times\underline{X})
=\underline{W}~.~(\underline{X}\times\underline{V})
=\underline{X}~.~(\underline{V}\times\underline{W})
\end{eqnarray}
If any two of $\underline{V}$, $\underline{W}$, $\underline{X}$ are the same then $\varepsilon^{abc}~V_a~W_b~X_c=0$.

Finally, 
\begin{eqnarray}
\frac{dT}{dt}
&=&\frac{1}{4\pi}~\varepsilon^{abc}\int\int\frac{\partial\phi_a}{\partial t}~\frac{\partial\phi_b}{\partial x}~\frac{\partial\phi_c}{\partial y}~dx~dy=\frac{1}{4\pi}\int\int 0~dx~dy\nonumber\\
&=&0
\end{eqnarray}
$T$ is a constant, no matter what non-linear sigma model we use, so long as
\begin{equation}
\phi_a
=\begin{pmatrix}
\sin f\sin g\\
\sin f\cos g\\
\cos f
\end{pmatrix}
\end{equation}

\section{HOPF CHARGE OF A TWISTED SKYRMION STRING}
Kobayashi and Nitta point out that the Hopf charge is defined to be \cite{nitta1}
\begin{eqnarray}
C
&=& \frac{1}{4\pi^2}\int dx^3~\varepsilon^{abc}~F_{ab}~A_c
\end{eqnarray}
where
\begin{eqnarray}
F_{ab}
&=& \vec\phi~.~(\partial_a\vec\phi\times\partial_b\vec\phi)
\end{eqnarray}
the field strength $A_c$ is a vector field satisfying the condition
\begin{eqnarray}
F_{ab}
&=& \partial_aA_b -\partial_bA_a
\end{eqnarray}
and $\varepsilon^{abc}$ is the alternating symbol, with
\begin{eqnarray}
\varepsilon^{123}=\varepsilon^{231}=\varepsilon^{312}=1;~~~\varepsilon^{213}=\varepsilon^{132}=\varepsilon^{213}=-1
\end{eqnarray}
and all other components are zero. It can be shown (using field equations for $\vec\phi$) that $C$ is conserved, meaning that $\partial_tC=0$, no matter what the geometry of the solution is.

In the twisted Skyrmion string model we have
\begin{eqnarray}
\vec\phi
&=&
\begin{pmatrix}
\sin f(r)~\sin g(\theta,z) \\
\sin f(r)~\cos g(\theta,z) \\
\cos f(r)
\end{pmatrix}
\end{eqnarray}
where
\begin{eqnarray}
g(\theta,z)
&=& n\theta + mkz
\end{eqnarray}
and so in view of the Chain Rule
\begin{eqnarray}
\partial_a\vec\phi
&=&
\begin{pmatrix}
\cos f(r)~\sin g(\theta,z) \\
\cos f(r)~\cos g(\theta,z) \\
-\sin f(r)
\end{pmatrix}
\frac{df}{dr}~\partial_ar +
\begin{pmatrix}
\sin f(r)~\cos g(\theta,z) \\
-\sin f(r)~\sin g(\theta,z) \\
0
\end{pmatrix} n~\partial_a\theta +mk~\partial_az
\end{eqnarray}
Taking the cross product of $\partial_a\vec\phi$ with $\partial_b\vec\phi$ gives:
\begin{eqnarray}
\partial_a\vec\phi\times\partial_b\vec\phi
&=& -(f'~\sin f)[(\partial_ar)(n~\partial_b\theta+mk~\partial_bz)-(\partial_br)(n~\partial_a\theta+mk~\partial_az)]
\begin{pmatrix}
\sin f(r)~\sin g(\theta,z) \\
\sin f(r)~\cos g(\theta,z) \\
\cos f(r)
\end{pmatrix}\\
&=& -(f'\sin f)[(\partial_ar)(n~\partial_b\theta+mk~\partial_bz)-(\partial_br)(n~\partial_a\theta+mk~\partial_az)]\vec\phi
\end{eqnarray}
and so
\begin{eqnarray}
F_{ab}
&=& \vec\phi~.~(\partial_a\vec\phi\times \partial_b\vec\phi) \nonumber\\
&=& -(f'\sin f)[(\partial_ar)(n~\partial_b\theta+mk~\partial_bz)-(\partial_br)(n~\partial_a\theta+mk~\partial_az)]
\end{eqnarray}
as $\vec\phi~.~\vec\phi=1$.

We now use the identities
\begin{eqnarray}
r
&=& (x^2+y^2)^{1/2};~~~\theta = \arctan\frac{y}{x}
\end{eqnarray}
and
\begin{eqnarray}
\partial_ar
&=&r^{-1}(\delta_a^xx +\partial_a^yy);~~~\partial_a\theta = r^{-2}(\delta_a^yx-\partial_a^xy)
\end{eqnarray}
to write
\begin{eqnarray}
\partial_ar(n~\partial_b\theta +mk~\partial_bz)-\partial_br(n~\partial_a\theta +mk~\partial_az)
&=& r^{-1}(\delta_a^xx+\delta_a^yy)[nr^{-2}(\delta_b^yx-\delta_b^xy) +mk~\delta_b^z] \nonumber\\
&&-~r^{-1}(\delta_b^xx+\delta_b^yy)[nr^{-2}(\delta_a^yx-\delta_a^xy)+mk~\delta_a^z] \nonumber\\
&=& mkr^{-1}[(\delta_a^xx+\delta_a^yy)\delta_b^z-(\delta_b^xx+\delta_b^yy)\delta_a^z] \nonumber\\
&&+~nr^{-3}[(\delta_a^xx+\delta_a^yy)(\delta_b^yx-\delta_b^xy)-(\delta_b^xx+\delta_b^yy)(\delta_a^yx-\delta_a^xy)] \nonumber\\
&=& mkr^{-1}[(\delta_a^xx+\delta_a^yy)\delta_b^z-(\delta_b^xx+\delta_b^yy)\delta_a^z] \nonumber\\
&&+~nr^{-1}(\delta_a^x\delta_b^y-\delta_a^y\delta_b^x)
\end{eqnarray}
Hence
\begin{eqnarray}
F_{ab}
&=& -\frac{df}{dr}\sin f~mkr^{-1}[(\delta_a^xx+\delta_a^yy)\delta_b^z -(\delta_b^xx+\delta_b^yy)\delta_a^z] -\frac{df}{dr}\sin fnr^{-1}(\delta_a^x\delta_b^y-\delta_a^y\delta_b^x)\nonumber\\
&=& mk[\partial_a(\cos f)\delta_b^z -\partial_b(\cos f)\delta_a^z] -\frac{df}{dr}\sin f~nr^{-1}(\delta_a^x\delta_b^y-\delta_a^y\delta_b^x)
\end{eqnarray}

We need to find a vector field $A_c$ with the property that $F_{ab}=\partial_aA_b - \partial_b A_a$. It turns out that
\begin{eqnarray}
A_c
&=& mk(\cos f)\delta_c^z +nr^{-2}(1+\cos f)(\delta_c^yx-\delta_c^xy)
\end{eqnarray}
The first term on the right is obvious from the expression for $F_{ab}$. We add the second term on the right because if
\begin{eqnarray}
A_x
&=& -nK(r)y;~~~A_y=nK(r)x
\end{eqnarray}
then
\begin{eqnarray}
\partial_yA_x
&=& -nK-nK'y^2r^{-1};~~~\partial_xA_y=nK +nK'x^2r^{-1}
\end{eqnarray}
and the equation for $A_c$ becomes
\begin{eqnarray}
-\frac{df}{dr}\sin f~nr^{-1}
&=& F_{xy} = \partial_xA_y -\partial_yA_x = nK +nK'x^2r^{-1} +nK +nK'y^2r^{-1} =n(2K+K'r)
\end{eqnarray}
The unknown function $K(r)$ therefore satisfies the differential equation
\begin{eqnarray}
2K +K'r = r^{-1}(\cos f)'
\end{eqnarray}
where a prime $(')$ denotes $d/dr$. Multiplying this equation by $r$ gives
\begin{eqnarray}
(Kr^2)' = (\cos f)'
\end{eqnarray}
and so after integrating we get
\begin{eqnarray}
Kr^2
&=& \cos f +const. = 1 +\cos f
\end{eqnarray}
The integration constant here is set to 1, because one of the boundary conditions of $f$ is that $f(0)=\pi$, and so $\cos f(0)=-1$. $K$ is therefore bounded at $r=0$ (meaning that $\lim_{r\rightarrow 0}Kr^2=0$) only if
\begin{eqnarray}
-1 +const. = 0
\end{eqnarray}
So, we conclude that
\begin{eqnarray}
K
&=& r^{-2}(1+\cos f)
\end{eqnarray}
Combining our expressions for $F_{ab}$ and $A_c$ gives
\begin{eqnarray}
\varepsilon^{abc}F_{ab}A_c
&=& \varepsilon^{abc}\left\{mk[\partial_a(\cos f)\delta_b^z -\partial_b(\cos f)\delta_a^z] -\frac{df}{dr}(\sin f)nr^{-1}(\partial_a^x\partial_b^y-\partial_a^y\partial_b^x)\right\}\nonumber\\
&&\times~ [mk(\cos f)\delta_c^z+nr^{-2}(1+\cos f)(\delta_c^yx-\delta_c^xy)]\nonumber\\
&=& mnkr^{-2}(1+\cos f)\varepsilon^{abc}[\partial_a(\cos f)\delta_b^z-\partial_b(\cos f)\delta_a^z](\delta_c^yx-\delta_c^xy) \nonumber\\
&&-~mnk~\frac{df}{dr}(\sin f\cos f)r^{-1}\varepsilon^{abc}(\delta_a^x\delta_b^y-\delta_a^y\delta_b^x)\delta_c^z \nonumber\\
&=& -2mnkr^{-2}(1+\cos f)[x~\partial_x(\cos f)-y~\partial_y(\cos f)] -2mnk\frac{df}{dr}(\sin f~\cos f)r^{-1}\nonumber\\
&=& -2mnkr^{-2}(1+\cos f)(-r~\sin f)\frac{df}{dr} -2mnk\frac{df}{dr}(\sin f~\cos f)r^{-1}\nonumber\\
&=& 2mnk\frac{df}{dr}(\sin f)r^{-1}
\end{eqnarray}
as
\begin{eqnarray}
\varepsilon^{abc}~\delta_a^x~\delta_b^y~\delta_c^z
&=& -\varepsilon^{abc}~\delta_a^y~\delta_b^x~\delta_c^z = 1
\end{eqnarray}
and
\begin{eqnarray}
\varepsilon^{abc}~\delta_b^z~\delta_c^y 
&=& -\delta_x^a;~~~\varepsilon^{abc}~\delta_b^z~\delta_c^x = \delta_y^a
\end{eqnarray}
If we integrate over a 3-dimensional volume with $r$ ranging from 0 to $\infty$, $\theta$ from 0 to $2\pi$, and $z$ over a finite vertical distance $\Delta z$, the enclosed Hopf charge is
\begin{eqnarray}
C
&=& \frac{1}{4\pi^2}\int dx^3~\varepsilon^{abc}~F_{ab}~A_c = \frac{mnk}{2\pi^2}\int \frac{df}{dr}(\sin f)r^{-1}~r~dr~d\theta~dz \nonumber\\
&=& -\frac{mnk}{\pi}[\cos f(\infty)-\cos f(0)]\Delta z =-\frac{2mnk}{\pi}\Delta z
\end{eqnarray}
as
\begin{eqnarray}
f(\infty)
&=& 0;~~~f(0) = \pi
\end{eqnarray}

It is clear from this expression for $C$ that the Hopf charge is undefined if $\Delta z\rightarrow \infty$. However, the Hopf charge is finite for solutions that are compact in the $z$-direction (meaning that the vortex has a finite length $\Delta z$, and $\vec\phi(z+\Delta z)=\vec\phi(z)$ for all $z$). Since the topological charge $n$ is known also to be conserved, it follows that $mk$ is separately conserved whenever the Hopf charge is conserved. In the case of our string solutions, a slight variation of this argument could also be used to show that $mk$ is again conserved. But, the question of whether $mk$ is conserved is part of a much more general question of whether our self-gravitating string solutions are stable, and this is something well beyond where we are at the moment.

\begin{center}
\textbf{Acknowlegment}
\end{center}
This research is funded fully by Graduate Research Scholarship Universiti Brunei Darussalam (GRS UBD). This support is greatly appreciated.

\end{document}